\documentclass[twocolumn]{svjour3}
\usepackage{graphics}
\usepackage{amsmath,amssymb}

\journalname{Eur. Phys. J. C}

\begin{document}

\title{Possible pentaquarks with heavy quarks}

\author{Hongxia Huang \and Chengrong Deng \and Jialun Ping\thanks{Corresponding
 author: jlping@njnu.edu.cn} \and Fan Wang}

\institute{Hongxia Huang \and Jialun Ping \at Department of Physics and Jiangsu Key Laboratory for Numerical
Simulation of Large Scale Complex Systems, Nanjing Normal University, Nanjing 210023, P. R. China \and
Chengrong Deng \at School of Mathematics and Physics, Chongqing Jiaotong University, Chongqing 400074, P.R. China \and
Fan Wang \at Department of Physics, Nanjing University, Nanjing 210093, P.R. China}

\titlerunning{Possible pentaquarks with heavy quarks}
\authorrunning{Huang,Deng,Ping,Wang}

\maketitle

\begin{abstract}
Inspired by the discovery of two pentaquarks $P_{c}(4380)$ and $P_{c}(4450)$ at the LHCb detector,
we study possible hidden-charm molecular pentaquarks in the framework of quark delocalization color
screening model. Our results suggest that both $N\eta_{c}$ with $IJ^{P}=\frac{1}{2}\frac{1}{2}^{-}$
and $NJ/\psi$ with $IJ^{P}=\frac{1}{2}\frac{3}{2}^{-}$ are bounded by channels coupling. However,
$NJ/\psi$ with $IJ^{P}=\frac{1}{2}\frac{3}{2}^{-}$ may be a resonance state in the $D-$wave $N\eta_{c}$
scattering process. Moreover, $P_{c}(4380)$ can be explained as the molecular pentaquark of
$\Sigma^{*}_{c}D$ with quantum numbers $IJ^{P}=\frac{1}{2}\frac{3}{2}^{-}$. The state $\Sigma^{*}_{c}D^{*}$
with $IJ^{P}=\frac{1}{2}\frac{5}{2}^{-}$ is a resonance, it may not be a good candidate of the observed
$P_{c}(4450)$ because of the opposite parity of the state to $P_{c}(4380)$, although the mass of the
state is not far from the experimental value. In addition, the calculation is extended to the hidden-bottom
pentaquarks, the similar properties as that of hidden-charm pentaquarks system are obtained.
\end{abstract}

\section{\label{sec:introduction}Introduction}

Multiquark states were studied even before the advent of quantum chromodynamics (QCD). The development
of QCD accelerated multiquark study because it is natural in QCD that there should be multiquark
states, glueballs and quark-gluon hybrids. After more than 40 years of quark model study,
the idea about baryon and meson is about to go beyond the naive picture: $q^3$ baryon and
$q\bar{q}$ meson. The proton spin puzzle could be explained by introducing the $q^3q\bar{q}$ component in
the quark model \cite{qcw}. In order to understand the baryon spectroscopy better, the five-quark
component of proton was proposed \cite{BSZou}. The baryon resonance is certainly coupled to the
meson-baryon scattering state and should be studied by coupling the $q^3$ with $q^3$-$q\bar{q}$
scattering channel in a quark model approach. Although the strange pentaquark state $\Theta^+$ claimed
by experimental groups thirteen years ago might be questionable (LEPS collaboration insists on the
existence of pentaquark $\Theta^+$~\cite{NakanoNew}) and the multi-quark states might be hard to
be identified, the multi-quark study is indispensable for understanding the low energy QCD,
because the multi-quark states can provide information unavailable in $q\bar{q}$ meson and $q^3$
baryon, especially the property of hidden color structure.

In the past decade, many near-threshold charmonium-like states have been observed at Belle,
BaBar, BESIII, and LHCb, triggering lots of studies on the molecule-like hadrons containing heavy quarks.
In the heavy quark sector, the large masses of the heavy quarks reduce the kinetic of the system,
which makes it easier to form bound states or resonances. So the heavy quarks play an important role to
stabilize the multiquark systems. There were many theoretical studies of hidden-charm
pentaquarks~\cite{Zou,YangZC,Uchino,Karliner}, especially the prediction of narrow $N^{*}$ and
$\Lambda^{*}$ resonances with hidden charm above 4 GeV by using the coupled-channel unitary
approach~\cite{Zou}, and the systematical investigation of possible hidden-charm molecular baryons
with components of an anti-charmed meson and a charmed baryon within the one boson exchange model~\cite{YangZC}.

Very recently, the LHCb Collaboration observed two pentaquark-charmonium states in the $J/\psi p$ invariant
mass spectrum of $\Lambda^{0}_{b} \rightarrow J/\psi K^{-}p$~\cite{LHCb}. One is $P_{c}(4380)$ with a mass of
$4380\pm8\pm29$ MeV and a width of $205\pm18\pm86$ MeV, and another is $P_{c}(4450)$ with a mass of
$4449.8\pm1.7\pm2.5$ MeV and a width of $39\pm5\pm19$ MeV. The preferred $J^{P}$ assignments are of opposite
parity, with one state having spin $\frac{3}{2}$ and the other $\frac{5}{2}$. Then, a lot of theoretical work
have been done to explain these two states. In Ref.~\cite{ChenHX}, the current experimental progress and
theoretical interpretations of the states were reviewed. R. Chen $et~al.$~\cite{ChenR} interpreted these two
hidden-charm states as the loosely bound $\Sigma_{c}(2455)D^{*}$ and $\Sigma^{*}_{c}(2520)D^{*}$ molecular
states by using the boson exchange model, and gave the spin parity $J^{P}=\frac{3}{2}^{-}$ and
$\frac{5}{2}^{-}$, respectively. While in Ref.~\cite{HeJ}, a Bethe-Salpeter equation approach was used to
studied the $\bar{D}\Sigma^{*}_{c}$ and $\bar{D}^{*}\Sigma_{c}$ interactions, and then $P_{c}(4380)$ and
$P_{c}(4450)$ were identified as $\bar{D}\Sigma^{*}_{c}$ and $\bar{D}^{*}\Sigma_{c}$ molecular states with
the spin parity $J^{P}=\frac{3}{2}^{-}$ and $\frac{5}{2}^{+}$, respectively. A QCD sum rule investigation
was performed, by which the $P_{c}(4380)$ was suggested as a $\bar{D}^{*}\Sigma_{c}$ hidden-charm pentaquark
with $J^{P}=\frac{3}{2}^{-}$ and the
$P_{c}(4450)$ was proposed as a mixed hidden-charm pentaquark of $\bar{D}^{*}\Lambda_{c}$ and $\bar{D}^{*}\Sigma_{c}$
with $J^{P}=\frac{5}{2}^{+}$~\cite{ChenHX2}. Also a coupled-channel calculation was performed to analyze the
$\Lambda^{0}_{b} \rightarrow J/\psi K^{-}p$ reaction and gave support to a $J^{P}=\frac{3}{2}^{-}$ assignment to
the $P_{c}(4450)$ and to its nature as a molecular state mostly made of $\bar{D}^{*}\Sigma_{c}$ and
$\bar{D}^{*}\Sigma^{*}_{c}$~\cite{Roca}. In Ref.~\cite{Oller}, Mei{\ss}ner and Oller suggested that the $P_{c}(4450)$ was almost entirely a $\chi_{c1}p$ resonance, coupling much more strongly to this channel than to $J/\psi p$. Kubarovsky and Voloshin~\cite{Kubarovsky} showed that the observed $P_{c}$ resonances are composites of $J/\psi$ and excited nucleon states with the quantum numbers of $N(1440)$ and $N(1520)$ within a simple "baryocharmonium" model. Moreover, some people proposed various rescattering mechanisms to
show that the $P_{c}(4450)$ state might arise from the kinematical effect~\cite{Guo,Liuxh}. Besides, Burns~\cite{Burns} explored the phenomenology of the $P_{c}(4380)$ and $P_{c}(4450)$ states, and their possible partners. Several intriguing similarities were also discussed in Ref.~\cite{Burns}, which suggested that the $P_{c}(4450)$ was related to the $X(3872)$ meson. Thus, different models
may give different descriptions for the resonance
structures. Clearly the quark level study of these two pentaquark-charmonium states is interesting and necessary.

It is well known that the nuclear force (the interaction between nucleons) are qualitative similar to the
molecular force (the interaction between atoms). This molecular model of nuclear forces, quark delocalization color
screening model (QDCSM)~\cite{QDCSM0}, has been developed and extensively studied. In this model, quarks confined in
one nucleon are allowed to delocalize to a nearby nucleon and the confinement interaction between quarks in different
baryon orbits is modified to include a color screening factor. The latter is a model description of the hidden color
channel coupling effect~\cite{QDCSM1}. The delocalization parameter is determined by the dynamics of the interacting
quark system, thus allows the quark system to choose the most favorable configuration through its own dynamics in
a larger Hilbert space. The model gives a good description of nucleon-nucleon and hyperon-nucleon interactions and
the properties of deuteron~\cite{QDCSM2}. It is also employed to calculated the baryon-baryon scattering phase shifts
and the dibaryon candidates in the framework of the resonating group method (RGM)~\cite{QDCSM3,QDCSM4}.

In this work, the resonating-group method (RGM) is employed to study the possible hidden-charm molecular
pentaquarks in QDCSM, and the channel-coupling effect are considered. Extension to the bottom case is
straightforward and is also included in the present work. The
structure of this paper is as follows. After the introduction, we present a brief introduction of the quark model used
in section II. Section III devotes to the numerical results and discussions. The summary is shown in the last section.

\section{The quark delocalization color screening
model (QDCSM)}

The detail of QDCSM used in the present work can be found  in the
references~\cite{QDCSM0,QDCSM1,QDCSM2,QDCSM3,QDCSM4}. Here, we
just present the salient features of the model. The model
Hamiltonian is:
\begin{eqnarray}
H &=& \sum_{i=1}^6 \left(m_i+\frac{p_i^2}{2m_i}\right) -T_c
+\sum_{i<j} V_{ij}, \\
V_{ij} & = &  V^{G}(r_{ij})+V^{\chi}(r_{ij})+V^{C}(r_{ij}),  \nonumber \\
V^{G}(r_{ij})&=& \frac{1}{4}\alpha_{s} \boldsymbol{\lambda}_i \cdot
\boldsymbol{\lambda}_j
\left[ \frac{1}{r_{ij}}-\frac{\pi}{2}\left(\frac{1}{m_{i}^{2}}
 +\frac{1}{m_{j}^{2}}+\frac{4\boldsymbol{\sigma}_i\cdot
 \boldsymbol{\sigma}_j}{3m_{i}m_{j}}  \right) \right. \nonumber \\
& & \left. ~~~~~~~~~~~~~~~~\delta(r_{ij})-\frac{3}{4m_{i}m_{j}r^3_{ij}}S_{ij}\right],
\nonumber \\
V^{\chi}(r_{ij})&=& \frac{\alpha_{ch}}{3}
\frac{\Lambda^2}{\Lambda^2-m_{\chi}^2}m_\chi \left\{ \left[
Y(m_\chi r_{ij})- \frac{\Lambda^3}{m_{\chi}^3}Y(\Lambda r_{ij})
\right] \right. \nonumber \\
&&  \boldsymbol{\sigma}_i \cdot\boldsymbol{\sigma}_j
\left. +\left[ H(m_\chi r_{ij})-\frac{\Lambda^3}{m_\chi^3}
H(\Lambda r_{ij})\right] S_{ij} \right\} \nonumber \\
& & {\mathbf F}_i \cdot
{\mathbf F}_j, ~~~\chi=\pi,K,\eta \nonumber
\end{eqnarray}
\begin{eqnarray}
V^{C}(r_{ij})&=& -a_c {\mathbf \lambda}_i \cdot {\mathbf
\lambda}_j [f(r_{ij})+V_0], \nonumber
\\
 f(r_{ij}) & = &  \left\{ \begin{array}{ll}
 r_{ij}^2 &
 \qquad \mbox{if }i,j\mbox{ occur in the same} \\
  & \mbox{~~~~~~baryon orbit} \\
  \frac{1 - e^{-\mu_{ij} r_{ij}^2} }{\mu_{ij}} & \qquad
 \mbox{if }i,j\mbox{ occur in different} \\
   & \mbox{~~~~~~baryon orbits} \\
 \end{array} \right.
\nonumber \\
S_{ij} & = &  \frac{{\mathbf (\sigma}_i \cdot {\mathbf r}_{ij})
({\mathbf \sigma}_j \cdot {\mathbf
r}_{ij})}{r_{ij}^2}-\frac{1}{3}~{\mathbf \sigma}_i \cdot {\mathbf
\sigma}_j. \nonumber
\end{eqnarray}
Where $S_{ij}$ is quark tensor operator; $Y(x)$ and $H(x)$ are
standard Yukawa functions~\cite{Valcarce}; $T_c$ is the kinetic
energy of the center of mass; $\alpha_{ch}$ is the chiral coupling
constant; determined as usual from the $\pi$-nucleon coupling
constant; $\alpha_{s}$ is the quark-gluon coupling constant. In
order to cover the wide energy range from light to heavy
quarks one introduces an effective scale-dependent quark-gluon
coupling $\alpha_{s}(\mu)$\cite{Vijande}:
\begin{eqnarray}
\alpha_{s}(\mu) & = &
\frac{\alpha_{0}}{\ln(\frac{\mu^2+\mu_{0}^2}{\Lambda_{0}^2})},
\end{eqnarray}
where $\mu$ is the reduced mass of two interacting quarks. All other symbols have their usual meanings.
Here, a phenomenological color screening confinement potential is used, and $\mu_{ij}$ is the color
screening parameter. For the light-flavor quark system, it is determined by fitting the deuteron
properties, $NN$ scattering phase shifts, $N\Lambda$ and $N\Sigma$ scattering phase shifts,
respectively, with $\mu_{uu}=0.45$, $\mu_{us}=0.19$ and $\mu_{ss}=0.08$, satisfying the relation,
$\mu_{us}^{2}=\mu_{uu}*\mu_{ss}$. When extending to the heavy quark case, there is no experimental data
available, so we take it as a adjustable parameter. In the present work, we take
$\mu_{cc}=0.01\sim 0.0001~{\rm fm}^{-2}$ and $\mu_{uc}$ is obtained by the relation
$\mu^{2}_{uc}=\mu_{uu}*\mu_{cc}$. All other parameters are also taken from our previous
work~\cite{QDCSM4}, except for the charm and bottom quark masses $m_{c}$ and $m_{b}$, which are
fixed by a fitting to the masses of the charmed and bottom baryons and mesons. The values of
those parameters are listed in Table \ref{parameters}. The calculated masses of the charmed
and bottom baryons and mesons are shown in Table \ref{baryons}.
\begin{table}[ht]
\begin{center}
\caption{Model parameters:
$m_{\pi}=0.7~{\rm fm}^{-1}$, $m_{k}=2.51~{\rm fm}^{-1}$,
$m_{\eta}=2.77~{\rm fm}^{-1}$, $\Lambda_{\pi}=4.2~{\rm fm}^{-1}$,
$\Lambda_{k}=5.2~{\rm fm}^{-1}$, $\Lambda_{\eta}=5.2~{\rm
fm}^{-1}$, $\alpha_{ch}=0.027$.}
\begin{tabular}{ccccc} \hline
$b$ & $m_{s}$ & $m_{c}$ & $m_{b}$ & $ a_c$  \\
(fm) & (MeV) & (MeV) & (MeV) & (MeV\,fm$^{-2}$)  \\ \hline\noalign{\smallskip}
0.518  &  573 & 1700 &    5140  &  58.03 \\ \hline\noalign{\smallskip}
$V_{0}$ &  $\alpha_{0}$ &  $\Lambda_{0}$ & $u_{0}$ &  \\
(MeV) &   &  (fm$^{-1}$) & (MeV) &  \\\hline\noalign{\smallskip}
 -1.2883 &  0.5101 &  1.525 &   445.808  & \\
\hline
\end{tabular}
\end{center}
\label{parameters}
\end{table}

\begin{table}[ht]
\begin{center}
\caption{The calculated masses (in MeV) of the charm and bottom baryons and mesons
in QDCSM. Experimental values are taken from the Particle Data Group (PDG)~\cite{PDG}. }
\begin{tabular}{lccccccc} \hline\noalign{\smallskip}
 & $\Sigma_{c}$ & $\Sigma^{*}_{c}$ & $\Lambda_{c}$ & $\Xi^{*}_{c}$ &
  $\Xi_{c}$ & $\Xi^{\prime}_{c}$  \\  \hline\noalign{\smallskip}
 Exp. & 2455 & 2520 & 2286 & 2645 & 2467 & 2575  \\  \hline\noalign{\smallskip}
Model & 2378 & 2404 & 2200 & 2552 & 2464 & 2533  \\  \hline\noalign{\smallskip}
 & $\Omega_{c}$ & $\Omega^{*}_{c}$ & $D$ & $D^{*}$ & $D_{s}$ & $D^{*}_{s}$ \\ \hline\noalign{\smallskip}
 Exp. & 2695 & 2770 & 1864 & 2007 & 1968 & 2112 \\ \hline\noalign{\smallskip}
Model & 2698 & 2709 & 1890 & 1924 & 2105 & 2119  \\ \hline\noalign{\smallskip}
& $B$ & $B^{*}$ &  $\eta_{c}$ & $J/\psi$ & $\eta_{b}$ & $\Upsilon(1s)$  \\ \hline\noalign{\smallskip}
 Exp.& 2980 & 3096 & 9391 & 9460 & 5279 & 5325  \\ \hline\noalign{\smallskip}
Model & 3224 & 3227 & 10104 & 10104 & 5333 & 5344  \\ \hline\noalign{\smallskip}
& $\Sigma_{b}$ & $\Sigma^{*}_{b}$ & $\Lambda_{b}$ & $\Xi_{b}$ &
 $\Omega_{b}$ &  \\ \hline\noalign{\smallskip}
 Exp. & 5811 & 5832 & 5619 & 5791 & 6071 &  \\  \hline\noalign{\smallskip}
Model & 5808 & 5816 & 5618 & 5887 & 6130 &  \\  \hline
\end{tabular}
\label{baryons}
\end{center}
\end{table}

The quark delocalization in QDCSM is realized by specifying the single particle orbital
wave function of QDCSM as a linear combination of left and right Gaussians, the single
particle orbital wave functions used in the ordinary quark cluster model,
\begin{eqnarray}
\psi_{\alpha}(\mathbf{s}_i ,\epsilon) & = & \left(
\phi_{\alpha}(\mathbf{s}_i)
+ \epsilon \phi_{\alpha}(-\mathbf{s}_i)\right) /N(\epsilon), \nonumber \\
\psi_{\beta}(-\mathbf{s}_i ,\epsilon) & = &
\left(\phi_{\beta}(-\mathbf{s}_i)
+ \epsilon \phi_{\beta}(\mathbf{s}_i)\right) /N(\epsilon), \nonumber \\
N(\epsilon) & = & \sqrt{1+\epsilon^2+2\epsilon e^{-s_i^2/4b^2}}. \label{1q} \\
\phi_{\alpha}(\mathbf{s}_i) & = & \left( \frac{1}{\pi b^2}
\right)^{3/4}
   e^{-\frac{1}{2b^2} (\mathbf{r}_{\alpha} - \mathbf{s}_i/2)^2} \nonumber \\
\phi_{\beta}(-\mathbf{s}_i) & = & \left( \frac{1}{\pi b^2}
\right)^{3/4}
   e^{-\frac{1}{2b^2} (\mathbf{r}_{\beta} + \mathbf{s}_i/2)^2}. \nonumber
\end{eqnarray}
Here $\mathbf{s}_i$, $i=1,2,...,n$ are the generating coordinates, which are introduced to
expand the relative motion wavefunction~\cite{QDCSM1}. The mixing parameter $\epsilon(\mathbf{s}_i)$
is not an adjusted one but determined variationally by the dynamics of the multi-quark system itself.
This assumption allows the multi-quark system to choose its favorable configuration in the interacting
process. It has been used to explain the cross-over transition between hadron phase and
quark-gluon plasma phase~\cite{Xu}.

\section{The results and discussions}

\begin{figure*}
\begin{center}
\resizebox{0.8\textwidth}{!}{\includegraphics{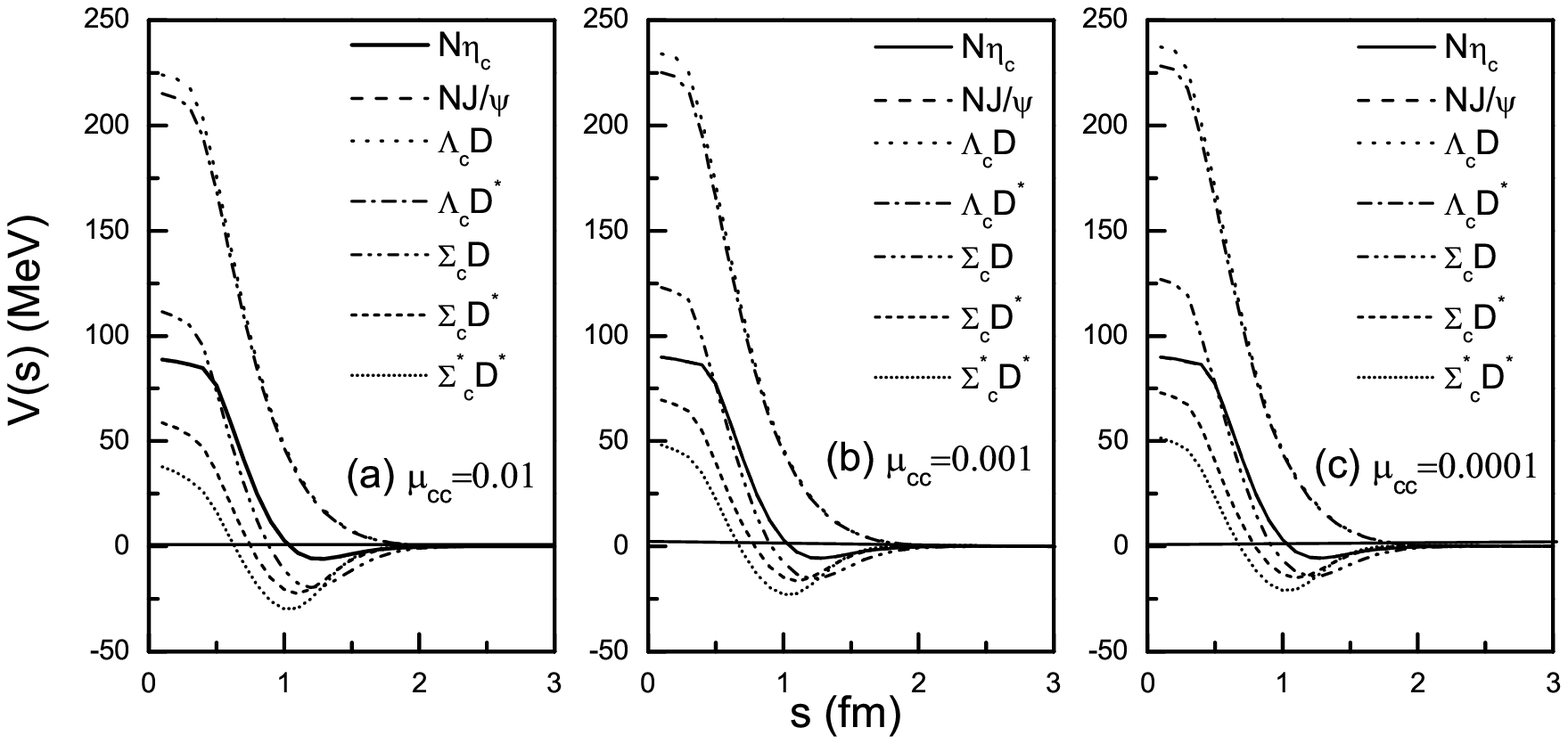}}

\caption{The potentials of different channels for the
$IJ^{P}=\frac{1}{2}\frac{1}{2}^{-}$ system.}

\resizebox{0.8\textwidth}{!}{\includegraphics{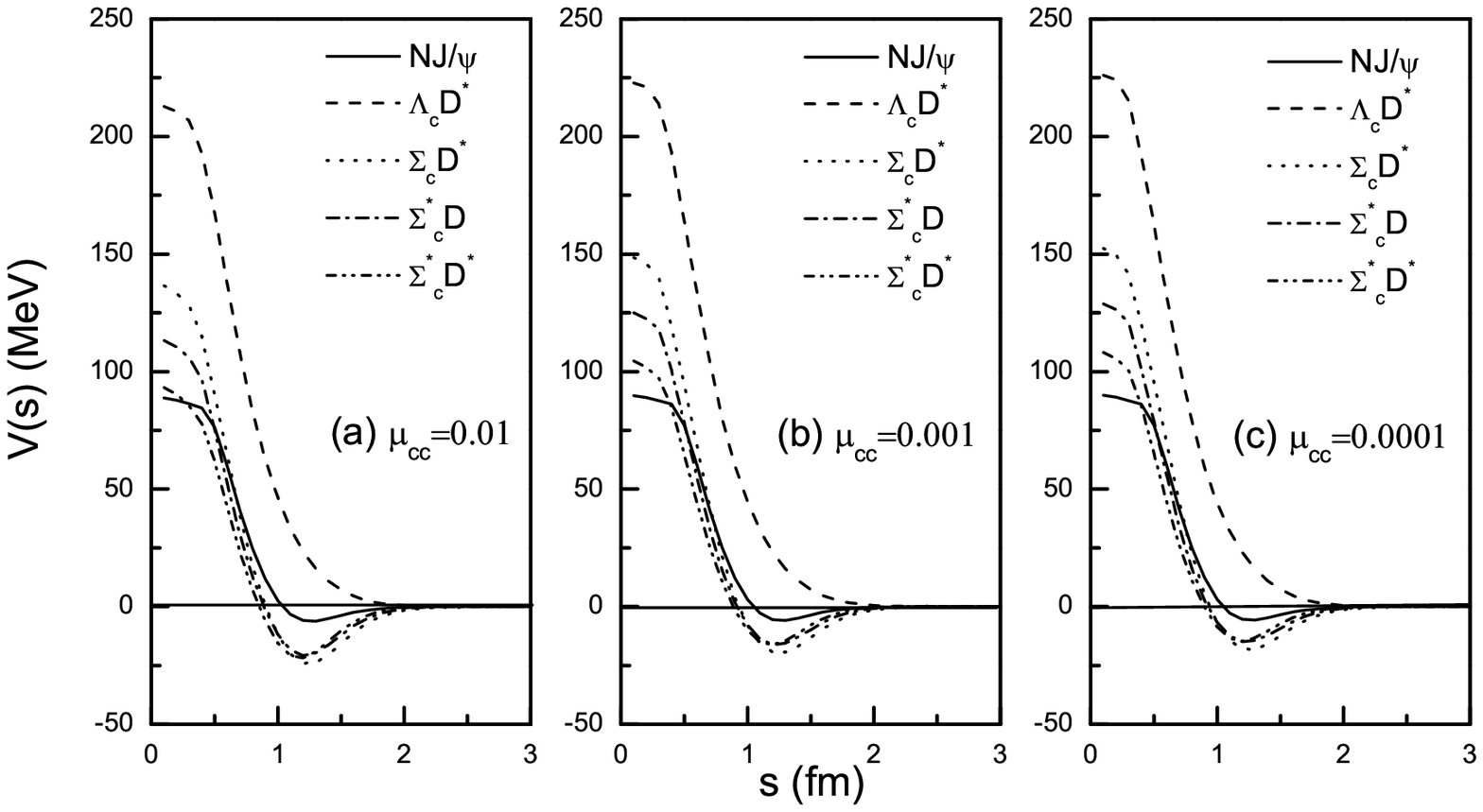}}

\caption{The potentials of different channels for the
$IJ^{P}=\frac{1}{2}\frac{3}{2}^{-}$ system.}
\end{center}
\end{figure*}

Here, we investigate the possible hidden-charm molecular pentaquarks with $Y=1$, $I=\frac{1}{2}$, $J^{P}=\frac{1}{2}^{\pm}$, $\frac{3}{2}^{\pm}$, and $\frac{5}{2}^{\pm}$. For the negative parity states, we calculate the $S$-wave channels with spin $S=\frac{1}{2}$, $\frac{3}{2}$, and $\frac{5}{2}$, respectively; and for the positive parity states, we calculate the $P$-wave channels with spin $S=\frac{1}{2}$, $\frac{3}{2}$, and $\frac{5}{2}$, respectively. All the channels involved are listed in Table~\ref{channels}. In the present calculation, we only consider the hidden-charm
molecular pentaquarks which consist of two $S$-wave hadrons. The channel coupling effects are also taken into account.
However, we find there is no any bound state with the positive parity within our calculations. There may exist other molecular structures, which contain excited hadrons, such as $\chi_{c1}p$ resonance~\cite{Oller}, $J/\psi N(1440)$, $J/\psi N(1520)$~\cite{Kubarovsky} and so on, which are out of range of present calculation. In the following
we only show the results of the negative parity states.

\begin{table}[h]
\caption{The channels involved in the calculation.}
\begin{center}
\begin{tabular}{lccccc}
\hline \noalign{\smallskip}
 $S=\frac{1}{2}$~~ & $N\eta_{c}$ & $NJ/\psi$  & $\Lambda_{c}D$ & $\Lambda_{c}D^{*}$
                  & $\Sigma_{c}D$ \\
    & $\Sigma_{c}D^{*}$  & $\Sigma^{*}_{c}D^{*}$   &¡¡& &  \\  \hline\noalign{\smallskip}
 $S=\frac{3}{2}$&  $NJ/\psi$ & $\Lambda_{c}D^{*}$
                  & $\Sigma_{c}D^{*}$ & $\Sigma^{*}_{c}D$
                  & $\Sigma^{*}_{c}D^{*}$ \\  \hline\noalign{\smallskip}
 $S=\frac{5}{2}$& $\Sigma^{*}_{c}D^{*}$ \\  \hline
\end{tabular}
\end{center}
\label{channels}
\end{table}

First, the effective potentials between two hadrons are calculated and shown in Figs. 1-3, because
an attractive potential is necessary for forming a bound state or resonance. The effective potential
between two colorless clusters is defined as, $V(s)=E(s)-E(\infty)$, where $E(s)$ is the energy of
the system at the separation $s$ of two clusters, which is obtained by the adiabatic approximation.
As mentioned in Sec. II, a phenomenological color screening confinement potential
is introduced in our model. For the multiquark systems with heavy quark, because no experimental data
is available, so we take three different values of $\mu_{cc}$ ($\mu_{cc}=0.01,~0.001,~0.0001$), to check
the dependence of our results on this parameter.
\begin{figure*}
\begin{center}
\resizebox{0.80\textwidth}{!}{\includegraphics{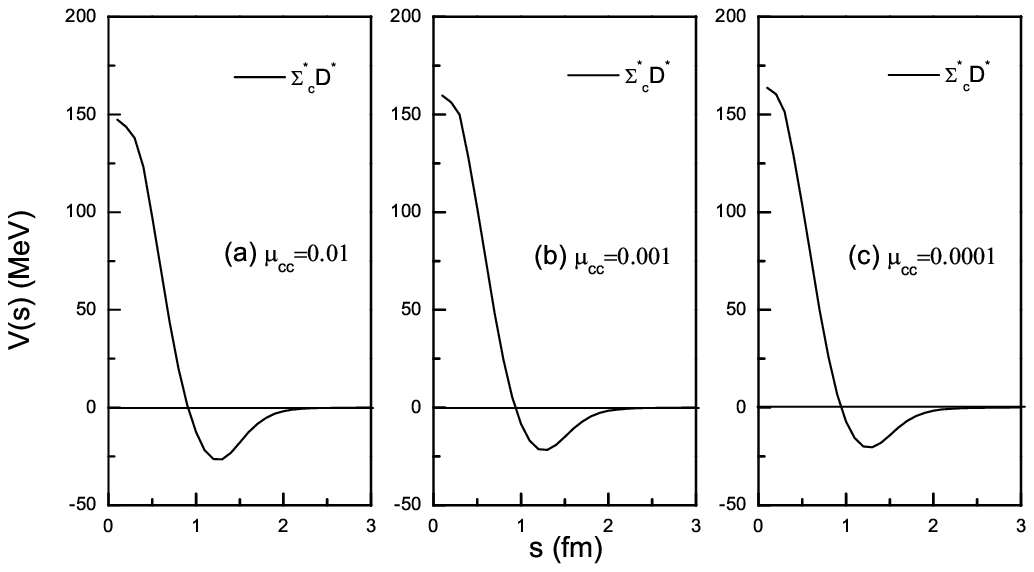}}

\caption{The potential of a single channel for the
$IJ^{P}=\frac{1}{2}\frac{5}{2}^{-}$ system.}
\end{center}
\end{figure*}

For the $J^{P}=\frac{1}{2}^{-}$ system (Fig. 1), one sees that the potentials are all
attractive for the channels $N\eta_{c}$, $NJ/\psi$, $\Sigma_{c}D$, $\Sigma_{c}D^{*}$ and
$\Sigma^{*}_{c}D^{*}$. While for the channels $\Lambda_{c}D$ and $\Lambda_{c}D^{*}$, the potentials
are repulsive, so no bound states or resonances can be formed in these two channels.
However, the bound states or resonances are possible for other channels due to the attraction nature
of the interaction between two hadrons.
The attraction between $\Sigma^{*}_{c}$ and $D^{*}$ is the largest one, followed by that of the
$\Sigma_{c}D^{*}$ channel, which is a little larger than that of the $\Sigma_{c}D$ channel.
In addition, the attraction of $N\eta_{c}$ is almost the same with that of $NJ/\psi$, which is the
smallest one during these five attractive channels.
Comparing figures (a), (b) and (c) in Fig. 1, we also find that larger values of
$\mu_{cc}$ give rise lower energy, although the variation is not very significant.

For the $J^{P}=\frac{3}{2}^{-}$ system (Fig. 2), similar results as that of
$IJ^{P}=\frac{1}{2}\frac{1}{2}^{-}$ system are obtained. The potentials are all
attractive for channels $NJ/\psi$, $\Sigma_{c}D^{*}$, $\Sigma^{*}_{c}D$ and $\Sigma^{*}_{c}D^{*}$,
while for the $\Lambda_{c}D^{*}$ channel, it is strongly repulsive. For the dependence
of potentials on the different values of $\mu_{cc}$, the behavior is the same as that for the
$J^{P}=\frac{1}{2}^{-}$ system.

For the $J^{P}=\frac{5}{2}^{-}$ system (Fig. 3), there is only one channel
$\Sigma^{*}_{c}D^{*}$, the potentials are attractive, and the dependence of potentials
on $\mu_{cc}$ are similar with that in
$J^{P}=\frac{1}{2}^{-}$ and $J^{P}=\frac{3}{2}^{-}$ system.

In order to see whether or not there is any bound states or resonances, a dynamic calculation is needed.
The resonating group method (RGM), described in more detail in Ref.\cite{Kamimura}, is used here.
Expanding the relative motion wavefunction between two clusters in the RGM by gaussians,
the integro-differential equation of RGM can be reduced to algebraic equation, the generalized
eigen-equation. The energy of the system can be obtained by solving the eigen-equation. In the
calculation, the baryon-meson separation ($|\mathbf{s}_n|$) is taken to be less than 6 fm
(to keep the matrix dimension manageably small).
\begin{table*}[ht]
\begin{center}
\caption{The binding energies and the masses (in MeV) of the hidden-charm molecular pentaquarks of $I=\frac{1}{2}$.}
{\begin{tabular}{cccc|cccc} \hline
 \multicolumn{4}{c|}{$J^{P}=\frac{1}{2}^{-}$} & \multicolumn{4}{c}{$J^{P}=\frac{3}{2}^{-}$} \\ \hline
 $\mu_{cc}$ & ~~{$0.01$} & ~~{$0.001$} & ~~{$0.0001$}~~ & $\mu_{cc}$  &~~{$0.01$} & ~~{$0.001$} & {$0.0001$} \\ \hline
 {$N\eta_{c}$}  &~~ ub & ~~ub & ~~~ub~~ & {$NJ/\psi$} & ~~ub & ~~ub  & ~~ub~~  \\
 {$NJ/\psi$}  &~~ ub & ~~ub & ~~~ub~~ & {$\Lambda_{c}D^{*}$} & ~~ub & ~~ub  & ~~ub~~  \\
 {$\Lambda_{c}D$}  &~~ ub & ~~ub & ~~~ub~~ & {$\Sigma_{c}D^{*}$} &  -16/4303 & ~-11/4308  & -10/4309  \\
 {$\Lambda_{c}D^{*}$}  &~~ ub & ~~ub & ~~~ub~~ & {$\Sigma^{*}_{c}D$} &  -17/4445 & -14/4448  & -12/4450  \\
   $\Sigma_{c}D$   &  -19/4300 &  -15/4304 & -13/4306 & {$\Sigma^{*}_{c}D^{*}$} &  -17/4510 & -15/4512  & -13/4514  \\ \cline{5-8}
   $\Sigma_{c}D^{*}$   &  -21/4441  &  -19/4443    &   -18/4444 & \multicolumn{4}{c}{$J^{P}=\frac{5}{2}^{-}$}    \\ \cline{5-8}
   $\Sigma^{*}_{c}D^{*}$ &  -24/4503  &  -23/4504    &   -21/4506  & {$\Sigma^{*}_{c}D^{*}$} & -15/4512 & -10/4517  & -10/4517 \\
\hline
\end{tabular}
\label{bound_c}}
\end{center}
\end{table*}

For the $J^{P}=\frac{1}{2}^{-}$ system, the single channel calculation shows that both
$\Lambda_{c}D$ and $\Lambda_{c}D^{*}$ are unbound, which agree with the repulsive nature of the
interaction of these two channels.
For the $N\eta_{c}$ and $NJ/\psi$ channels, the attractions are too weak to tie the two particles together,
the calculation shows that they are also unbound.
While, due to the stronger attractions, the obtained lowest energies of $\Sigma_{c}D$, $\Sigma_{c}D^{*}$
and $\Sigma^{*}_{c}D^{*}$ are below their corresponding thresholds.
The binding energy of these three states are listed in Table \ref{bound_c}, in which '$ub$' means
unbound. Here we should mention how we obtain the mass of a hidden-charm molecular pentaquark.
Generally, the mass of a molecular pentaquark can be written as $M^{the.}=M^{the.}_{1}+M^{the.}_{2}+B$,
where $M^{the.}_{1}$ and $M^{the.}_{2}$ stand for the theoretical masses of a charmed baryon and
an anti-charmed meson respectively, and $B$ is the binding energy of this molecular state.
In order to minimize the theoretical errors and to compare calculated results to the experimental
data, we shift the mass of molecular pentaquark to $M=M^{exp.}_{1}+M^{exp.}_{2}+B$, where the experimental
values of charmed baryons and anti-charmed mesons are used. Taking the state
$J^P={\frac12}^{-}~\Sigma_{c}D$ as an example, the calculated mass of pentaquark is
$4249$ MeV, then the binding energy $B$ is obtained by subtracting the theoretical masses of
$\Sigma_c$ and $D$, $4249-2378-1890=-19$ (MeV). Adding the experimental masses of the
hadrons, the mass of the pentaquark $M=2455+1864+(-19)=4300$ (MeV) is arrived.
In the present calculation, the resonance masses for $\Sigma_{c}D,~\Sigma_{c}D^{*}$ and
$\Sigma^{*}_{c}D^{*}$ with $J^{P}=\frac{1}{2}^{-}$ are $4300\sim4306$ MeV,
$4441\sim4444$ MeV, and $4503\sim4506$ MeV, respectively.
These results are qualitatively similar with the conclusion of Ref.\cite{Zou}, in which they
predicted two new $N^{*}$ states (the $\Sigma_{c}D$ molecular state $N^{*}(4265)$ and the
$\Sigma_{c}D^{*}$ molecular state $N^{*}(4415)$) in the coupled-channel unitary approach. Meanwhile, the chiral quark model calculation also supported the existence of the $S-$wave $\Sigma_{c}D$ bound state \cite{Wang}.
\begin{table*}[ht]
\begin{center}
\caption{The masses (in MeV) of the hidden-charm molecular pentaquarks of $J^{P}=\frac{1}{2}^{-}$ with three closed channels coupling
 and the percentages of each channel in the eigen-states.}
{\begin{tabular}{@{}cccc|ccc|ccc@{}c} \hline
  \multicolumn{1}{c}{$\mu_{cc}$}&\multicolumn{3}{c|}{$0.01$}
 &\multicolumn{3}{c}{$0.001$}&\multicolumn{3}{|c}{$0.0001$}\\ \hline
     {$M_{cc}$} &~~ {4296} & ~~{4437} & ~~{4500} ~~&~~ {4300} & ~~{4439} & ~~{4501} ~~&~~ {4302}& ~~{4440}& ~~{4503}  \\ \hline
   $\Sigma_{c}D$   & ~~95.5  &   2.9    &  4.8    &  ~~96.0  &  2.5  &  4.5  &  ~~96.7  &  2.1  &  4.2 \\
   $\Sigma_{c}D^{*}$   & ~~3.6   &   95.1    &   0.8   & ~~ 3.2  &  95.3  &  1.0  & ~~ 2.7  &  95.7  &  1.1 \\
   $\Sigma^{*}_{c}D^{*}$ & ~~0.9   &   2.0    &   94.4   & ~~ 0.8  &  2.2  &  94.5  & ~~ 0.6  &  2.2  &  94.7 \\
\hline
\end{tabular}
\label{ratio_c1}}

\vspace{4mm}

\caption{The masses (in MeV) of the hidden-charm molecular pentaquarks with all channels coupling
 and the percentages of each channel in the eigen-states.}
{\begin{tabular}{@{}cccc|cccc|cccc@{}c} \hline
 \multicolumn{1}{c}{}&\multicolumn{3}{c|}{$J^{P}=\frac{1}{2}^{-}$}&\multicolumn{1}{|c}{}
 &\multicolumn{3}{c|}{$J^{P}=\frac{3}{2}^{-}$} &\multicolumn{1}{|c}{} &\multicolumn{3}{c}{$J^{P}=\frac{5}{2}^{-}$}\\ \hline
 $\mu_{cc}$ & ~~{$0.01$} & ~~{$0.001$} & ~~{$0.0001$}~~ & $\mu_{cc}$  &~~{$0.01$} & ~~{$0.001$} & ~~{$0.0001$}~~ & $\mu_{cc}$  &~~{$0.01$} & ~~{$0.001$} & ~~{$0.0001$}~~ \\ \hline
 {$M_{cc}$}  &~~ {3881} & ~~{3883} & ~~{3884} ~~& {$M_{cc}$} & ~~~{3997} & ~~{3998} & ~~{3998}~~ & {$M_{cc}$} & ~~~{4512} & ~~{4517} & ~~{4517}~~  \\\hline
 {$N\eta_{c}$}  &~~ 41.7 & ~~49.7 & ~~35.2~~ &~~ {$NJ/\psi$} & ~~80.8 & ~~71.0  & ~~62.1~~  &~~ {$\Sigma^{*}_{c}D^{*}$} & ~~100.0 & ~~100.0  & ~~100.0~~\\
 {$NJ/\psi$}  &~~ 23.1 & ~~24.4 & ~~29.3~~ & ~~{$\Lambda_{c}D^{*}$} & ~~8.7 & ~~11.9  & ~~15.9~~ & & & & & \\
 {$\Lambda_{c}D$}  &~~ 14.6 & ~~11.7 & ~~14.5~~ & ~~{$\Sigma_{c}D^{*}$} & ~~1.2 & ~~1.9  & ~~2.6~~ & & & & & \\
 {$\Lambda_{c}D^{*}$}  &~~ 0.9 & ~~0.4 & ~~2.0~~ & ~~{$\Sigma^{*}_{c}D$} & ~~3.5 & ~~5.8  & ~~7.3~~  & & & & &\\
   $\Sigma_{c}D$   &  ~~ 0.1 & ~ 4.8 & ~~6.0~~ & ~~{$\Sigma^{*}_{c}D^{*}$} & ~~5.8 & ~~9.4  & ~~12.1~~  & & & & &\\
   $\Sigma_{c}D^{*}$   & ~~ 4.5  & ~ 6.4    & ~~12.4 ~~&~~ &  &   & & & & &     \\
   $\Sigma^{*}_{c}D^{*}$ & ~~ 15.1   & ~ 2.6  &   ~~0.6 ~~ &~~ &  &    & & & & &\\
\hline
\end{tabular}
\label{ratio_c2}}

\vspace{4mm}

\caption{The masses (in MeV) of the hidden-charm molecular pentaquarks of $J^{P}=\frac{3}{2}^{-}$ with three closed channels coupling
 and the percentages of each channel in the eigen-states.}
{\begin{tabular}{@{}cccc|ccc|ccc@{}c} \hline
 \multicolumn{1}{c}{$\mu_{cc}$}&\multicolumn{3}{c|}{$0.01$}
 &\multicolumn{3}{c}{$0.001$}&\multicolumn{3}{|c}{$0.0001$}\\ \hline
     {$M_{cc}$} &~~ {4362} & ~~{4445} & ~~{4551} ~~&~~ {4365} & ~~{4450} &~~ {4553} ~~&~~ {4368}&~~ {4451}&~~ {4554}  \\\hline
   $\Sigma_{c}D^{*}$   & ~~3.8  &   96.2    &  1.4    &  ~~1.6  &  98.0  &  1.0  &  ~~1.2  &  98.5  &  0.8 \\
   $\Sigma^{*}_{c}D$   & ~~91.0  &   2.8    &  4.0    &  ~~94.1  &  1.0  &  3.7  &  ~~95.5  &  0.7  &  3.0 \\
   $\Sigma^{*}_{c}D^{*}$ & ~~5.2  &   1.0    &  94.6    &  ~~4.3  &  1.0  &  95.3  &  ~~3.3  &  0.8  &  96.2 \\
\hline
\end{tabular}
\label{ratio_c3}}
\end{center}
\end{table*}

At the same time, we also do a channel-coupling calculation. In this work, two kinds of channel-coupling
are performed. The first one is the coupling of three closed channels ($\Sigma_{c}D$, $\Sigma_{c}D^{*}$
and $\Sigma^{*}_{c}D^{*}$). The results, the lowest three eigen-energies and the percentages of coupling
channels for the three eigen-states, are shown in Table \ref{ratio_c1}. Taking the results of $\mu_{cc}=0.01$
as an example, we can see that the main component of the lowest eigen-states is $\Sigma_{c}D$, $\sim$ 95.5\%,
and the the energy is pushed down a little, compared with the single-channel calculation,
4300 to 4296. The main component of the second lowest state is the $\Sigma_{c}D^{*}$ with the percentage
of $95.1\%$; and the main component of the third lowest state is the $\Sigma^{*}_{c}D^{*}$ with the
percentage of $94.4\%$. The three eigen-energies are all smaller than the thresholds of the corresponding
main channels, and are stable against the change of the baryon-meson separations. The large percentage of
the main component and the small change of energy infer that the channel-coupling is very weak.
However, these three closed channels can be coupled to other four open channels, $N\eta_{c}$, $NJ/\psi$,
$\Lambda_{c}D$ and $\Lambda_{c}D^{*}$. The results of this channel-coupling calculation are shown in
Table \ref{ratio_c2}. We obtain a stable state, the mass of which is lower than the threshold of $N\eta_{c}$,
and the main component of this state is $N\eta_{c}$, with the percentage of $41.7\%$.
This shows that the $N\eta_{c}$ of $J^{P}=\frac{1}{2}^{-}$ is bounded by channel-coupling in our quark model
calculation, the energy is $3881\sim3884$ (MeV). In addition, we also obtain several quasi-stable states,
the masses of which are smaller than the thresholds of the corresponding main channels, but they fluctuate
around the eigen-energies obtained in the three closed channel coupling calculation. For example, the energy
of one quasi-stable state is 4296 MeV, it fluctuates around this energy with 2 MeV with the variation of
the baryon-meson separation. To confirm whether the states of $\Sigma_{c}D$, $\Sigma_{c}D^{*}$ and
$\Sigma^{*}_{c}D^{*}$ can survive as resonance states after the full channel coupling, the study of the
scattering processes of the open channels of $N\eta_{c}$, $NJ/\psi$, $\Lambda_{c}D$ and $\Lambda_{c}D^{*}$
is needed. This work is underway. From the fluctuation, we can estimate the partial decay widths of these states
to $N\eta_{c}$, $NJ/\psi$, $\Lambda_{c}D$ and $\Lambda_{c}D^{*}$ are around several MeVs, if they are
resonances.

For the $J^{P}=\frac{3}{2}^{-}$ system, the similar results with the case of $J^{P}=\frac{1}{2}^{-}$
system are obtained. The single channel calculation shows that $NJ/\psi$ and $\Lambda_{c}D^{*}$ are
unbound, while $\Sigma_{c}D^{*}$, $\Sigma^{*}_{c}D$ and $\Sigma^{*}_{c}D^{*}$ are all bound.
The results are also listed in Table \ref{bound_c}. These three states also exist when they are
coupled together, the masses and the percentages of each channel of the lowest three eigen-states
are shown in Table \ref{ratio_c3}. We can see that the mass of the first eigen-state is about
$4362\sim4368$ MeV and the main channel is $\Sigma^{*}_{c}D$ with the percentage of $91.0\%\sim95.5\%$;
the mass of the second eigen-state is about $4445\sim4451$ MeV and the main channel is $\Sigma_{c}D^{*}$
with the percentage of $96.2\%\sim98.5\%$; the mass of the third eigen-state is about $4551\sim4555$ MeV
and the main channel is $\Sigma^{*}_{c}D^{*}$ with the percentage of $94.6\%\sim96.2\%$.
From the above results, we find that the mass of the first eigen-state is close to the mass
of the observed $P_{c}(4380)$, a pentaquark reported by LHCb collaboration. Therefore, in our quark model
calculation the main component of the $P_{c}(4380)$ is $\Sigma^{*}_{c}D$ with the quantum number
$J^{P}=\frac{3}{2}^{-}$. In addition, the mass of the second eigen-state is close to the mass of another
reported pentaquark $P_{c}(4450)$. Nevertheless, the opposite parity of the state to $P_{c}(4380)$ may
prevent this assignment. Moreover, all these closed channels can be coupled to the open channels $NJ/\psi$
and $\Lambda_{c}D^{*}$. The results of these five channels coupling are shown in Table \ref{ratio_c2}.
There is a stable state, the mass of which is lower than the threshold of $NJ/\psi$, and the main channel
of this state is $NJ/\psi$, with the ratio of $80.8\%\sim62.1\%$. This shows that the $NJ/\psi$ of
$J^{P}=\frac{3}{2}^{-}$ is bounded by channel-coupling. However, it can couple to the $D-$wave
$N\eta_{c}$. So further work should be done to check whether the $J^{P}=\frac{3}{2}^{-}$ $NJ/\psi$ is
a resonance state in the $D-$wave $N\eta_{c}$ scattering process. In fact, the possible existence of a nuclear bound quarkonium state was proposed more than 20 years ago by Brodsky, Schmidt and de Teramond~\cite{Brodsky}; Gao, Lee, and Marinov~\cite{Gao} also predicted the existence of the $N\phi$ bound state, which is very similar with $NJ/\psi$ state; and the recent lattice QCD calculation also supported the existence of the strangenium-nucleus and the charmonium-nucleus bound states~\cite{Beane}. Therefore, searching for the $NJ/\psi$ resonance state is the interesting work in future.
In addition, there are also several quasi-stable states in the full channel-coupling calculation,
the masses of which are lower than the thresholds of the corresponding main channels and they
fluctuate several MeVs around their central values. It is just the behavior of a resonance. The amplitude
of the fluctuation can be taken as the decay width of the quasi-states. In quark model calculation,
the decay width of the $P_{c}(4380)$ candidate is too small to match the experimental value.
Further study is needed to check whether these eigen-states are resonance states in the $NJ/\psi$ and
$\Lambda_{c}D^{*}$ scattering process and to calculate the widths of other decay modes.

As mentioned above, by taking into account the channel-coupling effect, a bound state $N\eta_c$ is obtained 
for the $J^{P}=\frac{1}{2}^{-}$ system; and another bound state $NJ/\psi$ is obtained for the 
$J^{P}=\frac{3}{2}^{-}$ system. In these two systems, the coupling between the calculated $S-$wave channels 
is through the central force. In order to see the strength of these channel-coupling, we calculate 
the transition potentials of these channels. Here, we take the result of the $J^{P}=\frac{3}{2}^{-}$ system 
with $\mu_{cc}=0.01$ as an example. The transition potentials of five channels $NJ/\psi$, $\Lambda_{c}D^{*}$, 
$\Sigma_{c}D^{*}$, $\Sigma^{*}_{c}D$, and $\Sigma^{*}_{c}D^{*}$ are shown in Fig. 4. Obviously, it is a 
strong coupling among these channels that makes $NJ/\psi$ the bound state. The mechanism to form a bound
state has been proposed before. Eric S. Swanson proposed that the admixtures of $\rho J/\psi$ and 
$\omega J/\psi$ states were important for forming $X(3872)$ state~\cite{Swanson}, which was also demonstrated 
in Ref.~\cite{Fern} by T. Fern$\acute{a}$ndez-Caram$\acute{e}$s and collaborators. The mechanism also
applied to the study of $H$-dibaryon~\cite{QDCSM5}, in which the single channel $\Lambda\Lambda$ is unbound, 
but when coupled to the channels $N\Xi$ and $\Sigma\Sigma$, it becomes a bound state. The effect of 
channel-coupling of the $J^{P}=\frac{3}{2}^{-}$ system is the same as the one of the $J^{P}=\frac{1}{2}^{-}$ system.

\begin{figure}
\begin{center}
\resizebox{0.45\textwidth}{!}{\includegraphics{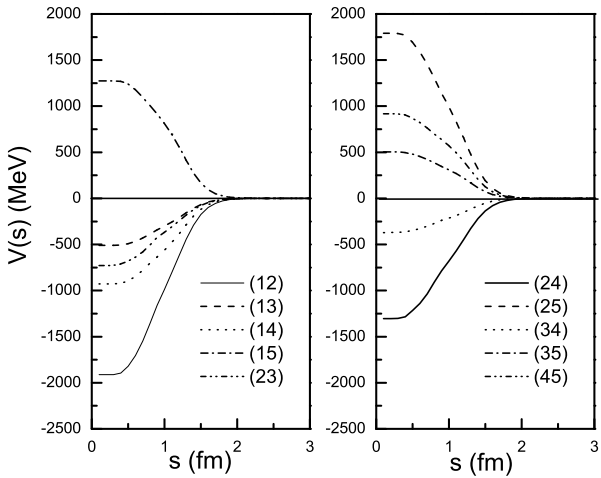}}

\caption{The transition potentials of channels $NJ/\psi$, $\Lambda_{c}D^{*}$, $\Sigma_{c}D^{*}$, $\Sigma^{*}_{c}D$, and $\Sigma^{*}_{c}D^{*}$, which are labeled as channels 1, 2, 3, 4, and 5 respectively.}
\end{center}
\end{figure}

For the $J^{P}=\frac{5}{2}^{-}$ system, it includes only one channel $\Sigma^{*}_{c}D^{*}$, and it is a
bound state with the mass of $4512\sim4517$ (MeV), which is a little higher than that of $P_{c}(4450)$.
Although the width of $S-$wave $\Sigma^{*}_{c}D^{*}$ decaying to $D-$wave $N\eta_{c}$ and $J/\psi p$
(tensor interaction induced decay) is generally small, which can be used to explain why the width of
$P_{c}(4450)$ is much narrower than that of $P_{c}(4380)$, the $J^{P}=\frac{5}{2}^{-}$ may not a good
candidate of $P_{c}(4450)$ because of the opposite parity of the state to $P_{c}(4380)$.

\begin{table}
\begin{center}
\caption{The binding energies (in MeV) of the hidden-bottom molecular pentaquarks of $I=\frac{1}{2}$.}
{\begin{tabular}{@{}cc|cc|cc@{}c} \hline
\multicolumn{2}{c|}{$J^{P}=\frac{1}{2}^{-}$}
 &\multicolumn{2}{c|}{$J^{P}=\frac{3}{2}^{-}$}&\multicolumn{2}{c}{$J^{P}=\frac{5}{2}^{-}$}\\ \hline
 {$N\eta_{b}$}  & ub &  {$N\Upsilon(1s)$} & ub   & {$\Sigma^{*}_{b}B^{*}$} & -14\\
 {$N\Upsilon(1s)$}  & ub  & {$\Lambda_{b}B^{*}$} & ub   \\
 {$\Lambda_{b}B$}  & ub  & {$\Sigma_{b}B^{*}$} &  -14   \\
 {$\Lambda_{b}B^{*}$}  & ub  & {$\Sigma^{*}_{b}B$} &  -15   \\
   $\Sigma_{b}B$   &   -15  & {$\Sigma^{*}_{b}B^{*}$} &  -16   \\
   $\Sigma_{b}B^{*}$   &  -21  & &  &  &  &      \\
   $\Sigma^{*}_{b}B^{*}$ &  -24     & &  &  &  &  \\
\hline
\end{tabular}
\label{bound_b}}
\end{center}
\end{table}

\begin{table}
\begin{center}
\caption{The masses (in MeV) of the hidden-bottom molecular pentaquarks with three closed channels coupling
 and the percentages of each channel in the eigen-states.}
{\begin{tabular}{ccccc} \hline
 & {$M_{cc}$} & {11070} & {11112} & {11132}  \\ \hline
 $J^{P}=\frac{1}{2}^{-}$  &  $\Sigma_{b}B$   & 76.8  &   12.4    &  8.1   \\
 &  $\Sigma_{b}B^{*}$   & 21.7  &   67.7    &  10.2    \\
 &  $\Sigma^{*}_{b}B^{*}$ & 1.5  &   19.9    &  81.7   \\ \hline
 & {$M_{cc}$} & {11091} & {11121} & {11138}   \\ \hline
 $J^{P}=\frac{3}{2}^{-}$  & {$\Sigma_{b}B^{*}$} &  5.1  &  86.5  &  9.5   \\
 &  $\Sigma^{*}_{b}B$ &  78.4  &  7.8  &  8.7   \\
 &  $\Sigma^{*}_{b}B^{*}$  &  16.5  &  5.7  &  81.8 \\ \hline
\end{tabular}
\label{ratio_b1}}
\end{center}
\end{table}

\begin{table}
\begin{center}
\caption{The masses (in MeV) of the hidden-bottom molecular pentaquarks of $I=\frac{1}{2}$
 and the percentages of each channel in the eigen-states.}
{\begin{tabular}{@{}cc|cc|cc@{}c} \hline
\multicolumn{2}{c|}{$J^{P}=\frac{1}{2}^{-}$}
 &\multicolumn{2}{c|}{$J^{P}=\frac{3}{2}^{-}$}&\multicolumn{2}{c}{$J^{P}=\frac{5}{2}^{-}$}\\ \hline
 {$M_{cc}$} & {10304} & {$M_{cc}$} & {10382} & {$M_{cc}$} & {11143}   \\ \hline
 {$N\eta_{b}$}  & 33.8 &  {$N\Upsilon(1s)$} & 34.6   & {$\Sigma^{*}_{b}B^{*}$} & 100.0\\
 {$N\Upsilon(1s)$}  & 14.7  & {$\Lambda_{b}B^{*}$} & 32.6   \\
 {$\Lambda_{b}B$}  & 24.2  & {$\Sigma_{b}B^{*}$} &  18.7   \\
 {$\Lambda_{b}B^{*}$}  & 5.2  & {$\Sigma^{*}_{b}B$} &  13.7   \\
   $\Sigma_{b}B$   &   2.1  & {$\Sigma^{*}_{b}B^{*}$} &  0.4   \\
   $\Sigma_{b}B^{*}$   &  0.7  & &  &  &  &      \\
   $\Sigma^{*}_{b}B^{*}$ &  19.3     & &  &  &  &  \\
\hline
\end{tabular}
\label{ratio_b2}}
\end{center}
\end{table}

In the previous discussion, the hidden-charm molecular pentaquarks were investigated. We also extend
the study to the hidden-bottom pentaquarks because of the heavy flavor symmetry. Here we take the
value of $\mu_{bb}=0.0001$. The numerical results are listed in Table \ref{bound_b}, Table \ref{ratio_b1}
and Table \ref{ratio_b2}. The results are similar to the hidden-charm molecular pentaquarks.
For the $J^{P}=\frac{1}{2}^{-}$ system, a bound state is obtained by all channels coupling, and the main
channel is $N\eta_{b}$ with the mass of $10304$ MeV; the quasi-stable states with main components of
$\Sigma_{b}B$, $\Sigma_{b}B^{*}$ and $\Sigma^{*}_{b}B^{*}$, respectively should be confirmed by calculating
the open channels scattering in future.
For the $J^{P}=\frac{3}{2}^{-}$ system, there is also a bound state of $10382$ MeV, and the main channel
is $N\Upsilon(1s)$, which also should be checked whether it is a resonance state or not in the
$D-$wave $N\eta_{b}$ scattering process. Moreover, further work should be done to check whether
the quasi-stable states of $\Sigma_{b}B^{*}$, $\Sigma^{*}_{b}B$ and $\Sigma^{*}_{b}B^{*}$ are resonance
states or not in the $N\Upsilon(1s)$ and $\Lambda_{b}B^{*}$ scattering process.
For the $J^{P}=\frac{5}{2}^{-}$ system, a bound state $\Sigma^{*}_{b}B^{*}$ is obtained, with the mass of $11143$ MeV.

\section{Summary}

In summary, the possible hidden-charm molecular pentaquarks with $Y=1$, $I=\frac{1}{2}$, $J^{P}=\frac{1}{2}^{\pm}$,
$\frac{3}{2}^{\pm}$, and $\frac{5}{2}^{\pm}$ are investigated by solving the RGM equation in the framework of QDCSM.
Our results show:
(1) All the positive parity states are all unbound in our calculation. Some other molecular structures, which contain excited hadrons, such as $\chi_{c1}p$, $J/\psi N(1440)$, $J/\psi N(1520)$ and so on, would deserve further study.
(2) For the $J^{P}=\frac{1}{2}^{-}$ system, there is a bound state of $3881\sim3884$ MeV by seven channels coupling, and the main channel
is $N\eta_{c}$; there are three quasi-stable states of $\Sigma_{c}D$, $\Sigma_{c}D^{*}$ and $\Sigma^{*}_{c}D^{*}$
should be confirmed by investigating the scattering process of the open channels of $N\eta_{c}$, $NJ/\psi$,
$\Lambda_{c}D$ and $\Lambda_{c}D^{*}$.
(3) For the $J^{P}=\frac{3}{2}^{-}$ system, the main channel of the bound state is $NJ/\psi$ with the mass of
$3997\sim3998$ MeV, which may be a resonance state in the $D-$wave $N\eta_{c}$ scattering process.
There are also three quasi-stable states: $\Sigma_{c}D^{*}$ with the mass of $4362\sim4368$ MeV,
$\Sigma^{*}_{c}D$ with the mass of $4445\sim4451$ MeV, and $\Sigma^{*}_{c}D^{*}$ with the mass of $4551\sim4555$ MeV,
of which the mass of $\Sigma_{c}D^{*}$ is close to the observed $P_{c}(4380)$. So in our quark model calculation
$P_{c}(4380)$ can be explained as the molecular pentaquark $\Sigma^{*}_{c}D$ with the quantum number
$J^{P}=\frac{3}{2}^{-}$. However, the partial decay width of $\Sigma_{c}D^{*}$ to $NJ/\psi$ is estimated to
 be several MeVs, which should be checked by further experiments. Similarly, the open channels of $NJ/\psi$ and $\Lambda_{c}D^{*}$ scattering process
calculation is needed to confirm the resonance states of $\Sigma_{c}D^{*}$, $\Sigma^{*}_{c}D$ and
$\Sigma^{*}_{c}D^{*}$.
(4) For the $J^{P}=\frac{5}{2}^{-}$ system, there is a bound state $\Sigma^{*}_{c}D^{*}$ with the mass of
$4512\sim4517$ (MeV). However, it may not a good candidate of the observed $P_{c}(4450)$ because of the opposite
parity of the state to $P_{c}(4380)$.
Besides, the calculation is also extended to the hidden-bottom pentaquarks. The results
are similar to the case of the hidden-charm molecular pentaquarks.

QDCSM, which was developed to study the multiquark states, is an extension of the na\"{i}ve quark model.
As we know, quark model plays an important role in the development of hadron physics.
The discovery of $\Omega^{-}$ is based on the prediction of quark concept of Gell-Mann-Zweig.
The na\"{i}ve quark model of Isgur {\em et~al.} gave a remarkable description of the properties
of ground-state hadrons. Applying to the excited states of hadron, hadron-hadron interaction and
multiquark systems, extensions to the na\"{i}ve quark model have to be made. Based on the different
extension of the naive quark model, a proliferation of bound states or resonances are predicted.
The recent progresses of experiments on $"XYZ"$ particles, $P_{c}^{+}$ pentaquarks and dibaryons
such as $d*$~\cite{dstar}, are encouraging.
However, some precaution about the proliferation of quark-model bound states has to be posed.
So far, there is no multiquark state identified by experiments unambiguously. For particular
multiquark state, there exist different points of view. For example, J. Vijande and collaborators
studied the four-quark system $c\bar{c}n\bar{n}$ in the constituent quark model by using different
types of quark-quark potentials, and no four-quark bound states have been found~\cite{Vijande1}, whereas
diquark-antidiquark picture was used by Maiani {\em et al.} to explain the state $X(3872)$ \cite{Maiani}.
J. Vijande {\em et~al.} also searched for the doubly-heavy dibaryons in a simple quark model,
but no bound or metastable state was found~\cite{Vijande2}, whereas $H$-like dibaryons with heavy quarks
were proposed in ref.\cite{Hlike}.
More theoretical and experimental work are needed to distinguish the different extension of the quark model.
The critical development of the quark model may be the unquenching quark model, where the valence quarks
and real/virtual quark pair are treated equally. T. F. Caram$\acute{e}$s and A. Valcarce have studied
the possible multiquark contributions to the charm baryon spectrum by considering higher order Fork space
components~\cite{Caram}. By incorporating new ingredients, the phenomenological quark model is expected to
describe ordinary and exotic hadrons well.

\section*{Acknowledgment}
The work is supported partly by the National Natural Science Foundation of China under Grant
Nos. 11175088, 11535005 and 11205091.


\begin{thebibliography}{99}
\bibitem{qcw} D. Qing, X. S. Chen and F. Wang, Phys. Rev. {\bf C57}, R31 (1998);
{\bf D58}, 114032 (1998).
\bibitem{BSZou} B. S. Zou and D. O. Riska, Phys. Rev. Lett. {\bf
 95}, 072001 (2005).
\bibitem{NakanoNew} T. Nakano, {\em et al.} (LEPS Collaboration), Phys. Rev. C. {\bf 79}, 025210 (2009).
\bibitem{Zou} J. J. Wu, R. Molina, E. Oset, and B. S. Zou, Phys. Rev. Lett. {\bf 105}, 232001
(2010); Phys. Rev. C. {\bf 84}, 015202 (2011).
\bibitem{YangZC} Z. C. Yang, Z. F. Sun, J. He, X. Liu and S. L. Zhu, Chin. Phys. C {\bf 36}, 6 (2012).
\bibitem{Uchino} T. Uchino, W. H. Liang and E. Oset, Eur. Phys. J. A. {\bf 52}, 43 (2016).
\bibitem{Karliner} M. Karliner and J. L. Rosner, Phys. Rev. Lett. {\bf 115}, 122001 (2015).
\bibitem{LHCb} R. Aaij, {\em et al.} (LHCb Collaboration), Phys. Rev. Lett. {\bf 115}, 072001 (2015).
\bibitem{ChenHX} H. X. Chen, W. Chen, X. Liu and S. L. Zhu, Phys. Rep. {\bf 639}, 1 (2016).
\bibitem{ChenR} R. Chen, X. Liu, X. Q. Li and S. L. Zhu, Phys. Rev. Lett. {\bf 115}, 132001 (2015).
\bibitem{HeJ} J. He, Phys. Lett. B. {\bf 753}, 547 (2016).
\bibitem{ChenHX2} H. X. Chen, W. Chen, X. Liu, T. G. Steel and S. L. Zhu, Phys. Rev. Lett. {\bf 115}, 172001 (2015).
\bibitem{Roca} L. Roca, J. Nieves and E. Oset, Phys. Rev. D. {\bf 92}, 094003 (2015).
\bibitem{Oller} Ulf-G. Mei{\ss}ner and J. A. Oller, Phys. Lett. B. {\bf 751}, 59 (2015).
\bibitem{Kubarovsky} V. Kubarovsky and M. B. Voloshin, Phys. Rev. D. {\bf 92}, 031502 (2015).
\bibitem{Guo} F. K. Guo, Ulf-G. Mei{\ss}ner, W. Wang and Z. Yang, Phys. Rev. D {\bf
92}, 071502 (2015).
\bibitem{Liuxh} X. H. Liu, Q. Wang and Q. Zhao, Phys. Lett. B. {\bf 757}, 231 (2016).
\bibitem{Burns} T. J. Burns, Eur. Phys. J. A. {\bf 51}, 152 (2015).
\bibitem{QDCSM0} F. Wang, G. H. Wu, L. J. Teng and T. Goldman, Phys. Rev. Lett. {\bf 69}, 2901 (1992);
G. H. Wu, L. J. Teng, J. L. Ping, F. Wang and T. Goldman, Phys. Rev. C {\bf
53}, 1161 (1996).
\bibitem{QDCSM1} H. X. Huang, P. Xu, J. L. Ping and F.
Wang, Phys. Rev. C {\bf 84}, 064001 (2011).
\bibitem{QDCSM2} J. L. Ping, F. Wang and T.
Goldman, Nucl. Phys. A {\bf 657}, 95 (1999); G. H. Wu, J. L. Ping,
L. J. Teng {\em et al.}, Nucl. Phys. A {\bf 673}, 279 (2000); H.
R. Pang, J. L. Ping, F. Wang and T. Goldman, Phys. Rev. C {\bf
65}, 014003 (2001); J. L. Ping, F. Wang and T. Goldman, Nucl.
Phys. A {\bf 688}, 871 (2001); J. L. Ping, H. R. Pang, F. Wang and
T. Goldman, Phys. Rev. C {\bf 65}, 044003 (2002).
\bibitem{QDCSM3} L. Z. Chen, H. R. Pang, H. X. Huang, J.
L. Ping and F. Wang, Phys. Rev. C {\bf 76}, 014001 (2007);  J. L.
Ping, H. X. Huang, H. R. Pang, F. Wang and C. W. Wong, Phys. Rev.
C {\bf 79}, 024001 (2009).
\bibitem{QDCSM4} M. Chen, H. X. Huang, J. L. Ping and F. Wang, Phys. Rev. C {\bf 83}, 015202 (2011).
\bibitem{Valcarce} A. Valcarce, H. Garcilazo, F. Fern\'{a}ndez and P. Gonzalez, Rep. prog. Phys.
{\bf 68}, 965 (2005) and reference there in.
\bibitem{Vijande} J. Vijande, F. Fernandez and A. Valcarce, J.
Phys. G {\bf 31}, 481 (2005).
\bibitem{PDG} J. Beringer, {\em et al.}, Particle Data Group, Phys. Rev. D {\bf 86}, 010001 (2012).
\bibitem{Xu} M. M. Xu, M. Yu and L. S. Liu, Phys. Rev. Lett. {\bf 100},
  092301 (2008).
\bibitem{Kamimura} M. Kamimura, Supp. Prog. Theo. Phys. {\bf 62}, 236 (1977).
\bibitem{Wang} W. L. Wang, F. Hang, Z. Y. Zhang and B. S. Zou, Phys. Rev. C {\bf 84}, 015203 (2011).
\bibitem{Brodsky} S. J. Brodsky, I. Schmidt and G. F. de Teramond, Phys. Rev. Lett. {\bf 64}, 1011 (1990).
\bibitem{Gao} H. Gao, T. -S. H. Lee, and V. Marinov, Phys. Rev. C {\bf 63}, 022201(R) (2001).
\bibitem{Beane} S. R. Beane, E. Chang, S. D. Cohen, W. Detmold, H.-W. Lin, K. Orginos, A. Parreno and M. J. Savage, Phys. Rev. D {\bf 91}, 114503 (2015).
\bibitem{Swanson} Eric S. Swanson, Phys. Lett. B. {\bf 588}, 189 (2004).
\bibitem{Fern} T. Fern$\acute{a}$ndez-Caram$\acute{e}$s, A. Valcarce, and J. Vijande, Phys. Rev. Lett. {\bf 103}, 222001 (2009).
\bibitem{QDCSM5} H. R. Pang, J. L. Ping, F. Wang, T. Goldman and E. G. Zhao, Phys. Rev. C {\bf 69}, 065207 (2004).
\bibitem{dstar} M. Bashkanov {\em et al} (CELSIUS-WASA Collaboration), Phys. Rev. Lett. {\bf 102}, 052301 (2009);
P. Adlarson {\em et al} (WASA-at-COSY Collaboration), Phys. Rev. Lett. {\bf 106}, 242302 (2011); {\bf 112}, 202301 (2014).
\bibitem{Vijande1} J. Vijande, E. Weissman, N. Barnea, and A. Valcarce, Phys. Rev. D {\bf 76}, 094022 (2007).
\bibitem{Maiani} L. Maiani, F. Piccinini, A. D. Polosa, and V. Riquer, Phys. Rev. D {\bf 71}, 014028 (2005).
\bibitem{Vijande2} J. Vijande, A. Valcarce, J. -M. Richard, and P. Sorba, Phys. Rev. D {\bf 94}, 034038 (2016).
\bibitem{Hlike} H. X. Huang. J. L. Ping, and F. Wang, Phys. Rev. C {\bf 92}, 035201 (2014).
\bibitem{Caram} T. F. Caram$\acute{e}$s and A. Valcarce, Phys. Rev. D {\bf 90}, 014042 (2014).


\end{thebibliography}
\end{document}